# The viscosity of charged particles in the weakly ionized plasma with power-law distributions

Yue Wang and Jiulin Du [a]

*Department of Physics, School of Science, Tianjin University, Tianjin 300072, China*

**Abstract** We study the viscosity of light charged particles in weakly ionized plasma with the power-law $q$-distributions by using the Boltzmann equation of transport and the motion equation of hydrodynamics. The nonequilibrium plasma is considered to be space inhomogeneous and without magnetic field. We derive the expressions of the viscosity coefficients of the electrons and the ions in the $q$-distributed plasma, including the first and the second viscosity coefficients respectively. It is shown that these new viscosity coefficients depend strongly on the $q$-parameters, and when we take $q\rightarrow 1$, they perfectly return to those in the plasma with a Maxwellian distribution. The discussions presented in this paper can also be applied to the plasmas with the kappa-distributions.

I. INTRODUCTION

The power-law distributions are a class of non-Maxwellian distributions or non-exponential distributions, typical examples such as the $q$-distributions in complex systems studied within the framework of nonextensive statistics,[1] the $\kappa$-distributions observed in astrophysical and space plasmas,[2-4] and the α-distributions, e.g., $E^{-\alpha}$, commonly existed in some nonequilibium systems in the fields of physics, chemistry, astronomy, biology and engineering technology etc.[5-10] Theoretically, the power-law distributions can be determined from a stochastic dynamics of the Brown's motion having a generalized fluctuation-dissipation relation between the friction coefficient and the diffusion coefficient.[11]

In plasmas physics, non-Maxwellian distributions have been observed and studied commonly both in astrophysical and space plasmas and in laboratory plasmas.[12] For instance, the plasmas in the planetary magnetospheres are nonequilibrium and found to be far deviation from the Maxwellian distribution due to the presence of high energy particles.[13] The spacecraft measurements of plasma velocity distributions, both in the solar wind and in the planetary magnetospheres and magnetosheaths, have revealed that non-Maxwellian distributions are quite common. In many situations the velocity distributions have a "suprathermal" power-law tail at high energies, which has been well modeled by the famous $\kappa$-distribution.[2-4]

In recent years, the researches on the complex plasmas with power-law distributions have attracted great interest because their many interesting applications are found in the wide fields of space plasma physics and astrophysics, and also because the power-law distributions observed in plasmas can be studied under the framework of nonextensive statistics.[14-28] And with the aid of the nonextensive kinetic theory, one can determine the expressions of the nonextensive $q$-parameter in the astrophysical and space plasmas[29] and thereby understand its physical meaning.

Nonextensive statistics is a new statistical theory to generalize the traditional Boltzmann-Gibbs statistics to that can be reasonably used to study complex systems, including the systems with long-range interactions, such as plasma and self-gravitating systems.[1] On the basis of the nonextensive *q*-entropy and the probabilistically independent postulate, the nonextensive statistics is found to have the so-called power-law *q*-distribution functions. The power-law velocity *q*-distribution function can be expressed [29,30] as

$$f_q^{(0)}(\mathbf{v}) = nB_q \left(\frac{m}{2\pi k_B T}\right)^{\frac{3}{2}} \left[1-(1-q)\frac{m(\mathbf{v}-\mathbf{u})^2}{2k_B T}\right]^{1/(1-q)}, \quad (1)$$

where *q* is a nonextensive parameter who's deviation from 1 represents the degree of nonextensivity, *T* is temperature, **u** is the bulk velocity of the fluid under consideration, *n* is the number density of particles, *m* is mass of the particle, $k_B$ is Boltzmann constant, and $B_q$ is the *q*-dependent normalized constant given by

$$B_q = \begin{cases} (1-q)^{\frac{1}{2}}(3-q)(5-3q)\dfrac{\Gamma\left(\frac{1}{2}+\frac{1}{1-q}\right)}{4\Gamma\left(\frac{1}{1-q}\right)}, & \text{for } q \leq 1. \\ (q-1)^{\frac{3}{2}}\dfrac{\Gamma\left(\frac{1}{q-1}\right)}{\Gamma\left(\frac{1}{q-1}-\frac{3}{2}\right)}, & \text{for } q \geq 1. \end{cases}$$

---
[a] jldu@tju.edu.cn

In general, the temperature *T*, the density *n* and the bulk velocity **u** in the nonequilibrium complex plasma should be considered to be space inhomogeneous, i.e., *T*=*T*(*r*), *n*=*n*(*r*) and **u**=**u**(*r*), and they can vary with time. The velocity *q*-distribution is not an equilibrium distribution but is a nonequilibrium stationary distribution. In 2004, the equation of the *q*-parameter was first derived for the nonequlibrium plasma with Coulombian long-range interactions,[29]

$$k_B \nabla T = (1-q)\, e \nabla \varphi,$$

where *e* is electron charge and *φ* is the Coulomb potential. This equation tell us clearly that we have *q*≠1 if and only if ∇*T*≠0. Therefore, the *q*-distribution describes the properties at the nonequiibrium stationary-state of the plasma in the external field, and only if we let *q*→1 it becomes the Maxwell distribution at a thermal equilibrium-state. The plasmas under consideration are the complex plasmas obeying the non-extensive statistics. The power-law velocity *q*-distribution has been extensively applied to deal with varieties of problems in the systems with long-range interactions, which can describe the statistical property of the nonequilibrium complex plasmas with the electromagnetic interactions.

The transport coefficients in the power-law distributed plasma were first studied under a very simplified Lorentz model.[31] Most recently, the diffusion coefficients in the weakly ionized plasma with the κ-distribution [32] and some transport properties of the full ionized plasma with the κ-distribution [33] were studied. As we know, the weakly ionized plasma contains less charged particle than neutral particle, and the interactions between particles are taking place by the collisions, such as electron-electron collision, electron-ion collision, electron-neutral-particle collision and ion-neutral-particle collision. Since the number of charged particles is small, when we calculate the transport coefficients we mainly consider the collisions of charged particles with the neutral particles and can neglect the collisions between the charged particles.

In the plasma, if the mean velocity of a certain plasma component is space inhomogeneous, there are the momentum currents to move from a high velocity layer to a low velocity layer, leading the viscosity. As important one of transport processes in plasma, the viscosity describes the momentum transport. In this work, we will study the viscosity in the weakly ionized plasma with the power-law $q$-distributions and then derive the new viscosity coefficients in this situation.

The paper is organized as follows. In Sec.II, we introduce the transport equations for the weakly ionized plasma with the power-law $q$-distributions in nonextensive statistics. In Sec.III, we study the viscosity of charged particles in the $q$-distributed plasma and derive the viscosity coefficients in this physical situation, including the first viscosity coefficient and the second viscosity coefficient respectively. Finally in Sec.IV, we give the conclusion.

## II. TRANSPORT EQUATION

The Boltzmann form of the collision integral together with further simplifications can be used to study the transport properties of weakly ionized plasmas. The near collisions between the ionized plasma particles and the neutral background gas are the main mechanisms for determining the transport coefficients. As usual, for the weakly ionized plasma without magnetic field, the collisions between charged particles and neutral molecules or atoms are generally treated as an elastic collision if the plasma chemistry is not considered, and the inelastic collisions are neglected in the transport theory.[34-37] However, the duration of such a collision is generally much smaller than the times between successive collisions. This fact together with the molecular chaos hypothesis constitutes the basis of the statistical model of a low-density plasma gas, and the mathematical representation of this model is the Boltzmann transport equation. In nonextensive statistics, the statistical description of this multi-body system can be given by the generalized Boltzmann equation,[38]

$$\frac{\partial f_\alpha}{\partial t} + \mathbf{v} \cdot \frac{\partial f_\alpha}{\partial \mathbf{r}} + \frac{\mathbf{F}_\alpha}{m_\alpha} \cdot \frac{\partial f_\alpha}{\partial \mathbf{v}} = C_q(f_\alpha), \qquad (2)$$

where $f_\alpha \equiv f_\alpha(\mathbf{r},\mathbf{v},t)$ is a single-particle velocity distribution function at time $t$, velocity $\mathbf{v}$, and position $\mathbf{r}$, the subscript $\alpha$ denotes the electron and the ion, $\alpha = e, i$, respectively, and $\mathbf{F}_\alpha$ is the external field force. In the following we appoint that $\mathbf{r} = (r_x, r_y, r_z) \equiv (x, y, z)$. The term on the right-hand side $C_q$ is the nonextensive $q$-collision term, which represents the change in $f_\alpha$ due to the collisions. As a generalization, the $q$-collision term should (can) contain those collision terms given in the traditional statistics, for different physical situations. It is verified [38] that the velocity $q$-distribution is a stationary solution of Eq.(2), and if the $q$-H theorem is satisfied, the $q$-collision term $C_q$ vanishes and $f_\alpha(\mathbf{r},\mathbf{v},t)$ will evolve irreversibly towards the velocity $q$-distribution.

The dynamical origin of the $q$-distribution in nonextensive statistics is a very important problem under investigation. Now it is believed that in the complex system there is a new fluctuation-dissipation relation between the friction and diffusion.[11] Under this condition, the long-time behavior of such a system does not approach to a Maxwell distribution but necessarily approach to the power-law velocity $q$-distribution.[39]

Following the line in textbooks, in the weakly ionized plasma, because the mass of neutral particles are heavy as compared with electrons and with ions if ions are seeded in the plasma, the neutral particles can be regarded as static and homogeneous distribution, and because the characteristic timescale for the charged particle collisions is much more than that for the charged-neutral collisions, we mainly consider the collisions between charged and neutral particles, and neglect the collision between the charged particles. In this way, here the Krook collision model can be generalized as [34]

$$C_q(f_\alpha) = -\frac{1}{\tau_\alpha}\left(f_\alpha - f_{q,\alpha}^{(0)}\right) = -\nu_\alpha\left(f_\alpha - f_{q,\alpha}^{(0)}\right), \tag{3}$$

where $\tau_\alpha$ (v) is the mean time of collisions between the charged and neutral particle and here it can be assumed to be a constant. Or if $\nu_\alpha$ is the mean collision frequency, we have that $\nu_\alpha = (\tau_\alpha)^{-1}$.

As usual, in the first-order approximation of Chapman-Enskog expansion, we can write the velocity distribution function as the following form:

$$f_\alpha = f_{q,\alpha}^{(0)} + f_{q,\alpha}^{(1)}, \tag{4}$$

where $f_{q,\alpha}^{(1)}$ is a first-order small disturbance about the stationary $q$-distribution $f_{q,\alpha}^{(0)}$. In the case of smooth flow plasma without magnetic field, according to the power-law velocity $q$-distribution (1), we have that for the $\alpha$th component,

$$f_{q,\alpha}^{(0)}(\mathbf{r},\mathbf{v}) = n_\alpha B_{q,\alpha}\left(\frac{m_\alpha}{2\pi k_B T_\alpha}\right)^{3/2}\left[1-(1-q_\alpha)\frac{m_\alpha}{2k_B T_\alpha}(\mathbf{v}-\mathbf{u})^2\right]^{1/(1-q_\alpha)}, \tag{5}$$

with

$$B_{q,\alpha} = \begin{cases} (1-q_\alpha)^{\frac{1}{2}}(3-q_\alpha)(5-3q_\alpha)\dfrac{\Gamma\left(\frac{1}{2}+\frac{1}{1-q_\alpha}\right)}{4\Gamma\left(\frac{1}{1-q_\alpha}\right)}, & \text{for } 0<q_\alpha\leq 1. \\[1em] (q_\alpha-1)^{\frac{3}{2}}\dfrac{\Gamma\left(\frac{1}{q_\alpha-1}\right)}{\Gamma\left(\frac{1}{q_\alpha-1}-\frac{3}{2}\right)}, & \text{for } q_\alpha\geq 1. \end{cases}$$

Substituting Eqs. (3)-(5) into Eq. (2), we obtain

$$\left(\frac{\partial}{\partial t}+\mathbf{v}\cdot\frac{\partial}{\partial \mathbf{r}}+\frac{\mathbf{F}_\alpha}{m_\alpha}\cdot\frac{\partial}{\partial \mathbf{v}}\right)\left(f_{q,\alpha}^{(0)}+f_{q,\alpha}^{(1)}\right) = -\nu_\alpha f_{q,\alpha}^{(1)}. \tag{6}$$

Transport processes are all studied in a steady state so that $\partial f_\alpha/\partial t = 0$. And the first-order small disturbance satisfies $f_{q,\alpha}^{(1)} \ll f_{q,\alpha}^{(0)}$, so we can neglect $f_{q,\alpha}^{(1)}$ on the left side of Eq.(6). Thus following the line in the textbooks,[34,35] we can obtain the first-order approximation expression of the distribution function,

$$f_{q,\alpha}^{(1)} = -\frac{1}{\nu_\alpha}\left[\mathbf{v}\cdot\frac{\partial}{\partial \mathbf{r}}+\frac{\mathbf{F}_\alpha}{m_\alpha}\cdot\frac{\partial}{\partial \mathbf{v}}\right]f_{q,\alpha}^{(0)}. \tag{7}$$

Therefore from Eq.(4) we write the distribution function,

$$f_\alpha = f_{q,\alpha}^{(0)} - \frac{1}{\nu_\alpha}\left[\mathbf{v}\cdot\frac{\partial}{\partial \mathbf{r}}+\frac{\mathbf{F}_\alpha}{m_\alpha}\cdot\frac{\partial}{\partial \mathbf{v}}\right]f_{q,\alpha}^{(0)}, \tag{8}$$

where the external field force is $\mathbf{F}_\alpha = Q_\alpha \mathbf{E}$ with the electric field intensity $\mathbf{E}$ and the electric charge $Q_\alpha$ for $\alpha$th component in the plasma.

## III. THE VISCOSITY COEFFICIENTS OF CHARGED PARTICLES

Transport processes in nonequilibrium plasma involve various thermodynamic "fluxes". These "fluxes" are driven by the corresponding thermodynamic "forces". Usually, if the "fluxes" are only considered the linearized part with respect to the "forces", the linear correlation coefficients are called as the transport coefficients. The viscosity phenomenon means that there are the momentum currents to move from a high velocity layer to a low velocity layer if the mean velocity of a certain plasma component is space inhomogeneous. Generally, according to the law of conservation of momentum, the motion equation of plasma hydrodynamics satisfies [36]

$$\frac{\partial(\rho \mathbf{u})}{\partial t} = -\nabla \cdot (\rho \mathbf{u}\mathbf{u} + \mathbf{P}) + \sum_\alpha \rho_\alpha \mathbf{f}_\alpha, \tag{9}$$

where $\rho \mathbf{u}$ is the momentum density, $(\rho \mathbf{u}\mathbf{u} + \mathbf{P})$ is the momentum current density tensor, $\rho$ is the total mass density, $\rho_\alpha$ is the mass density of $\alpha$th component, and $\rho_\alpha \mathbf{f}_\alpha$ is the external field forces (such as the gravity, and the electromagnetic forces) of $\alpha$th component of unit volume. The bulk velocity of fluid can be $\mathbf{u} = (u_x, u_y, u_z)$. $\mathbf{P}$ is the stress tensor, $\mathbf{P} = \{P_{ij}\}_{i,j=x,y,z}$, in which the matrix elements can be given[34,37] by

$$P_{ij} = p\delta_{ij} - \eta \left( \frac{\partial u_i}{\partial r_j} + \frac{\partial u_j}{\partial r_i} - \frac{2}{3} \frac{\partial u_k}{\partial r_k} \delta_{ij} \right) - \zeta \frac{\partial u_k}{\partial r_k} \delta_{ij}, \tag{10}$$

where $\eta$ is the first viscosity coefficient, $\zeta$ is the second viscosity coefficient or volume viscosity coefficient, $p$ is the static pressure, and according to the appointment we have that

$$\frac{\partial u_k}{\partial r_k} \equiv \sum_{k=x,y,z} \frac{\partial u_k}{\partial r_k}. \tag{11}$$

Eq. (9) combined with Eq.(10) is actually equal to Navier-Stokes equation.

On the other hand, if $f(\mathbf{r}, \mathbf{v}, t)$ is the velocity distribution function, the matrix elements of the stress tensor are defined [34,37] by

$$P_{ij} = \int m(v_i - u_i)(v_j - u_j) f(\mathbf{r}, \mathbf{v}, t) d\mathbf{v}. \tag{12}$$

Now, based on the above theory we can calculate the viscosity coefficients of the light charged particles in the weakly ionized plasma with the velocity $q$-distributions.

### A. The first viscosity coefficients

Using the above equation (10), we have that

$$P_{\alpha, xz} = -\eta_\alpha \left( \frac{\partial u_x}{\partial z} + \frac{\partial u_z}{\partial x} \right), \tag{13}$$

where $\eta_\alpha$ is the first viscosity coefficient for $\alpha$th component of the plasma. And according to Eq.(12), in our case the matrix element can be described by the integral,

$$P_{\alpha,xz} = \int m_\alpha (v_x - u_x)(v_z - u_z) f_\alpha(\mathbf{r},\mathbf{v}) d\mathbf{v}. \tag{14}$$

It represents the momentum component along the z-direction and across the unit area perpendicular to the x-axis per unit time. Using Eq.(8) we have that

$$P_{\alpha,xz} = \int d\mathbf{v}\, m_\alpha (v_x - u_x)(v_z - u_z) \left\{ f_{q,\alpha}^{(0)} - \frac{1}{\nu_\alpha} \left[ \mathbf{v} \cdot \frac{\partial}{\partial \mathbf{r}} + \frac{Q_\alpha \mathbf{E}}{m_\alpha} \cdot \frac{\partial}{\partial \mathbf{v}} \right] f_{q,\alpha}^{(0)} \right\}. \tag{15}$$

On the right-hand side in Eq.(15), the first term of the integral is given by

$$m_\alpha \int d\mathbf{v}(v_x - u_x)(v_z - u_z) f_{q,\alpha}^{(0)}$$

$$= m_\alpha n_\alpha B_{q,\alpha} \left( \frac{m_\alpha}{2\pi k_B T_\alpha} \right)^{3/2} \int d\mathbf{v}(v_x - u_x)(v_z - u_z) \left[ 1 - (1-q_\alpha) \frac{m_\alpha}{2k_B T_\alpha} (\mathbf{v}-\mathbf{u})^2 \right]^{1/(1-q_\alpha)}. \tag{16}$$

The integral in (16) is zero because the integrand is an odd function about ($v_x - u_x$) and ($v_z - u_z$). The third part of the integral in Eq.(15) is that

$$-\frac{Q_\alpha}{\nu_\alpha} \int d\mathbf{v}(v_x - u_x)(v_z - u_z) \left( \mathbf{E} \cdot \frac{\partial f_{q,\alpha}^{(0)}}{\partial \mathbf{v}} \right)$$

$$= \frac{2\pi Q_\alpha n_\alpha B_{q,\alpha}}{\nu_\alpha} \left( \frac{m_\alpha}{2\pi k_B T_\alpha} \right)^{5/2} \int d\mathbf{v}(v_x - u_x)(v_z - u_z) \mathbf{E} \cdot (\mathbf{v}-\mathbf{u}) \left[ 1 - (1-q_\alpha) \frac{m_\alpha (\mathbf{v}-\mathbf{u})^2}{2k_B T_\alpha} \right]^{\frac{1}{1-q_\alpha}-1}. \tag{17}$$

It is easy to find that the integrand is still an odd function about ($v_y - u_y$) and so the integral (17) for the velocity is equal to zero. Then Eq.(15) is reduced to

$$P_{\alpha,xz} = -\frac{m_\alpha}{\nu_\alpha} \int d\mathbf{v}(v_x - u_x)(v_z - u_z) \left[ \mathbf{v} \cdot \frac{\partial f_{q,\alpha}^{(0)}}{\partial \mathbf{r}} \right]. \tag{18}$$

Substituting the q-distribution function (5) into Eq.(18), it becomes

$$P_{\alpha,xz} = -\frac{2\pi m_\alpha n_\alpha B_{q,\alpha}}{\nu_\alpha} \left( \frac{m_\alpha}{2\pi k_B T_\alpha} \right)^{5/2} \int d\mathbf{v}(v_x - u_x)(v_z - u_z) A(\mathbf{r},\mathbf{v}) \left[ 1 - \frac{(1-q_\alpha)m_\alpha}{2k_B T_\alpha}(\mathbf{v}-\mathbf{u})^2 \right]^{\frac{1}{1-q_\alpha}-1}, \tag{19}$$

with

$$A(\mathbf{r},\mathbf{v}) = \sum_{k=x,y,z} (v_k - u_k) \mathbf{v} \cdot \frac{\partial u_k}{\partial \mathbf{r}}.$$

After the integral in Eq.(19) is calculated (see Appendix A), we derive that

$$P_{\alpha,xz} = -\frac{2n_\alpha k_B T_\alpha}{\nu_\alpha (7-5q_\alpha)} \left( \frac{\partial u_x}{\partial z} + \frac{\partial u_z}{\partial x} \right), \quad 0 < q_\alpha < \frac{7}{5}. \tag{20}$$

Comparing Eq.(20) with Eq.(13), we find the first viscosity coefficients of charged particles in the weakly ionized plasma with power-law q-distributions,

$$\eta_{q,\alpha} = \frac{2n_\alpha k_B T_\alpha}{\nu_\alpha (7-5q_\alpha)}, \quad 0 < q_\alpha < \frac{7}{5}, \tag{21}$$

We see that the viscosity coefficients depend strongly on the $q$-parameter in the power-law $q$-distributions. And when we take $q_\alpha \to 1$, Eq.(21) recovers the viscosity coefficient with a Maxwellian distribution,[34]

$$\eta_\alpha = \frac{n_\alpha k_B T_\alpha}{\nu_\alpha} \qquad (22)$$

**B. The second viscosity coefficients**

In plasma hydrodynamics, it is usually supposed that the arithmetic mean of the three diagonal elements of the stress tensor is equal to the sum of the pressure at the fixed point and the quantity that is proportional to the volume expansion rate $\nabla \cdot \mathbf{u}$. Namely,[34]

$$\frac{1}{3}\left(\mathbf{P}_{\alpha,xx} + \mathbf{P}_{\alpha,yy} + \mathbf{P}_{\alpha,zz}\right) = p_\alpha - \zeta_\alpha \nabla \cdot \mathbf{u}, \qquad (23)$$

where $\zeta_\alpha$ is the second viscosity coefficient for $\alpha$th component in the plasma. According to Eq.(12), in our case one of the diagonal elements can be described by the integral,

$$P_{\alpha,xx} = \int m_\alpha (v_x - u_x)^2 f_\alpha(\mathbf{r}, \mathbf{v}) d\mathbf{v}. \qquad (24)$$

Using Eq.(8) we can write Eq.(24) as

$$P_{\alpha,xx} = \int d\mathbf{v} m_\alpha (v_x - u_x)^2 \left\{ f_{q,\alpha}^{(0)} - \frac{1}{\nu_\alpha}\left[\mathbf{v} \cdot \frac{\partial}{\partial \mathbf{r}} + \frac{Q_\alpha \mathbf{E}}{m_\alpha} \cdot \frac{\partial}{\partial \mathbf{v}}\right] f_{q,\alpha}^{(0)} \right\}$$

$$= P_\alpha - \frac{m_\alpha}{\nu_\alpha} \int d\mathbf{v}(v_x - u_x)^2 \left[\mathbf{v} \cdot \frac{\partial f_{q,\alpha}^{(0)}}{\partial \mathbf{r}} + \frac{Q_\alpha \mathbf{E}}{m_\alpha} \cdot \frac{\partial f_{q,\alpha}^{(0)}}{\partial \mathbf{v}}\right], \qquad (25)$$

where the static pressure is defined [34] as

$$p_\alpha = m_\alpha \int d\mathbf{v}(v_x - u_x)^2 f_{q,\alpha}^{(0)}. \qquad (26)$$

So we have that

$$p_\alpha = m_\alpha n_\alpha B_{q,\alpha} \left(\frac{m_\alpha}{2\pi k_B T_\alpha}\right)^{\frac{3}{2}} \int d\mathbf{v}(v_x - u_x)^2 \left[1 - (1-q_\alpha)\frac{m_\alpha (\mathbf{v}-\mathbf{u})^2}{2k_B T_\alpha}\right]^{\frac{1}{1-q_\alpha}}$$

$$= \frac{2 n_\alpha k_B T_\alpha}{7 - 5 q_\alpha}, \quad 0 < q_\alpha < \frac{7}{5}. \qquad (27)$$

The third term in Eq.(25) is the integral given by

$$-\frac{Q_\alpha}{\nu_\alpha} \int d\mathbf{v}(v_x - u_x)^2 \mathbf{E} \cdot \frac{\partial f_{q,\alpha}^{(0)}}{\partial \mathbf{v}}$$

$$= \frac{2\pi Q_\alpha n_\alpha}{\nu_\alpha} B_{q,\alpha} \left(\frac{m_\alpha}{2\pi k_B T_\alpha}\right)^{\frac{5}{2}} \int d\mathbf{v}(v_x - u_x)^2 \mathbf{E} \cdot (\mathbf{v}-\mathbf{u}) \left[1 - \frac{(1-q_\alpha) m_\alpha (\mathbf{v}-\mathbf{u})^2}{2 k_B T_\alpha}\right]^{\frac{1}{1-q_\alpha}-1} = 0. \qquad (28)$$

The second term in Eq.(25) is the integral that

$$-\frac{m_\alpha}{\nu_\alpha}\int d\mathbf{v}(v_x-u_x)^2\,\mathbf{v}\cdot\frac{\partial f_{q,\alpha}^{(0)}}{\partial \mathbf{r}}$$

$$=-\frac{2\pi n_\alpha m_\alpha}{\nu_\alpha}B_{q,\alpha}\left(\frac{m_\alpha}{2\pi k_B T_\alpha}\right)^{\frac{5}{2}}\int d\mathbf{v}(v_x-u_x)^2 A(\mathbf{r},\mathbf{v})\left[1-\frac{(1-q_\alpha)m_\alpha}{2k_B T_\alpha}(\mathbf{v}-\mathbf{u})^2\right]^{\frac{1}{1-q_\alpha}-1},$$

$$=-\frac{2n_\alpha k_B T_\alpha}{\nu_\alpha(7-5q_\alpha)}\left(3\frac{\partial u_x}{\partial x}+\frac{\partial u_y}{\partial y}+\frac{\partial u_z}{\partial z}\right),\quad 0<q_\alpha<\frac{7}{5}, \tag{29}$$

where $A(\mathbf{r},\mathbf{v})$ is the same as that in Eq.(19). The calculations for the integral in the second equation of Eq.(29) can be seen in Appendix B.

Substituting Eqs.(27)-(29) into Eq.(25), we obtain that

$$P_{\alpha,xx}=\frac{2n_\alpha k_B T_\alpha}{7-5q_\alpha}\left[1-\frac{1}{\nu_\alpha}\left(3\frac{\partial u_x}{\partial x}+\frac{\partial u_y}{\partial y}+\frac{\partial u_z}{\partial z}\right)\right]. \tag{30}$$

In the same way as above, we can derive the other two diagonal elements of the stress tensor, respectively, represented by

$$P_{\alpha,yy}=\frac{2n_\alpha k_B T_\alpha}{7-5q_\alpha}\left[1-\frac{1}{\nu_\alpha}\left(\frac{\partial u_x}{\partial x}+3\frac{\partial u_y}{\partial y}+\frac{\partial u_z}{\partial z}\right)\right], \tag{31}$$

and

$$P_{\alpha,zz}=\frac{2n_\alpha k_B T_\alpha}{7-5q_\alpha}\left[1-\frac{1}{\nu_\alpha}\left(\frac{\partial u_x}{\partial x}+\frac{\partial u_y}{\partial y}+3\frac{\partial u_z}{\partial z}\right)\right]. \tag{32}$$

Substituting Eq.(27) and Eqs.(30)-(32) into Eq.(23), we find the second viscosity coefficients of charged particles in the weakly ionized plasma with power-law $q$-distributions,

$$\zeta_\alpha=\frac{10 n_\alpha k_B T_\alpha}{3\nu_\alpha(7-5q_\alpha)}. \tag{33}$$

We see that the viscosity coefficients also depend strongly on the $q$-parameter in the power-law $q$-distributions. And when we take $q_\alpha\to 1$, Eq.(33) also recovers the second viscosity coefficient with a Maxwellian distribution,

$$\zeta_\alpha=\frac{5 n_\alpha k_B T_\alpha}{3\nu_\alpha}. \tag{34}$$

IV. CONCLUSION AND DISCUSSION

In conclusion, we have studied the viscosity of charged particles in the weakly ionized plasma with power-law $q$-distributions in nonextensive statistics. By using the Boltzmann transport equation and the motion equation of hydrodynamics, we investigated the matrix elements of the stress tensor in the plasma with power-law $q$-distributions and therefore we derived the expressions of the viscosity coefficients of both the electrons and the ions, respectively including the first viscosity coefficients, given by Eq.(21), and the second viscosity coefficients or the volume viscosity coefficients, given by Eq.(33). These new viscosity coefficients depend strongly on the nonextensive $q$-parameter in the power-law $q$-distributions, and when we take the limit $q\to1$, they perfectly return to those in the case of a Maxwellian distribution.

In our results (21) and (33), we have the limitation of the value for the $q$-parameter, i.e. $0<q<7/5$. In fact, such limitation for the $q$-parameter exists generally in the state equation in nonextensive statistics if the average in velocity space is made using the standard definition.[40] Using the equation of the $q$-parameter, combined with the experimental data, one can determine the $q$-parameter values of a system. For example, in the solar interiors, all the values of the $q$-parameter are found always to be in the limitation range.[41]

The discussions of the viscosity coefficients for the power-law $q$-distributions in this paper may also be applied to the plasmas with the power-law $\kappa$-distributions as long as we make the parameter replacement by $(q-1)^{-1}=\kappa$,[14] or by $(q-1)^{-1}=\kappa+1$.[42] We can also follow the same ways in this paper to derive the results for the $\kappa$-distributed plasma by replacing the $q$-distribution functions with the $\kappa$-distribution ones.

*Additional remarks*: In Eq.(6), if an equation for the electric field force $F_\alpha \equiv Q_\alpha E = (k_B T_\alpha)/L$ was satisfied by the plasma, the Knudsen number $K_n \ll 1$ may be equivalent to the condition $f_{q,\alpha}^{(1)} \ll f_{q,\alpha}^{(0)}$ from Eq.(6) to Eq.(7) in the first-order approximation of the stationary Boltzmann equation, where $L$ is the characteristic scale of the plasma system, and $K_n = l/L$ with the average free path $l$ of the light charged particles.

**ACNOWLEDGEMENT**

This work is supported by the National Natural Science Foundation of China under Grant No. 11775156.

**APPENDIX A**

Eq.(19) is

$$P_{\alpha,xz} = -\frac{2\pi m_\alpha n_\alpha B_{q,\alpha}}{\nu_\alpha}\left(\frac{m_\alpha}{2\pi k_B T_\alpha}\right)^{\frac{5}{2}}\int d\mathbf{v}(v_x-u_x)(v_z-u_z)A(\mathbf{r},\mathbf{v})\left[1-\frac{(1-q_\alpha)m_\alpha}{2k_B T_\alpha}(\mathbf{v}-\mathbf{u})^2\right]^{\frac{1}{1-q_\alpha}-1} \quad (A.1)$$

where the integral is calculated, for $q_\alpha > 1$, by

$$\int d\mathbf{v}(v_x-u_x)(v_z-u_z)A(\mathbf{r},\mathbf{v})\left[1-\frac{(1-q_\alpha)m_\alpha}{2k_B T_\alpha}(\mathbf{v}-\mathbf{u})^2\right]^{\frac{1}{1-q_\alpha}-1}$$

$$=\int_{-\infty}^{+\infty}d\mathbf{v}(v_x-u_x)(v_z-u_z)\left[1-\frac{(1-q_\alpha)m_\alpha}{2k_B T_\alpha}(\mathbf{v}-\mathbf{u})^2\right]^{\frac{1}{1-q_\alpha}-1}\left[(v_x-u_x)v_z\frac{\partial u_x}{\partial z}+(v_z-u_z)v_x\frac{\partial u_z}{\partial x}\right]$$

$$=\left(\frac{\partial u_x}{\partial z}+\frac{\partial u_z}{\partial x}\right)\int_{-\infty}^{+\infty}d\mathbf{v}(v_x-u_x)^2(v_z-u_z)^2\left[1-\frac{(1-q_\alpha)m_\alpha}{2k_B T_\alpha}(\mathbf{v}-\mathbf{u})^2\right]^{\frac{1}{1-q_\alpha}-1}.$$

$$=\left(\frac{\partial u_x}{\partial z}+\frac{\partial u_z}{\partial x}\right)\int_0^{2\pi}d\varphi\int_0^\pi d\theta\int_0^{+\infty}dv\, v^6(\sin\theta\cos\theta\cos\varphi)^2\left[1-\frac{(1-q_\alpha)m_\alpha}{2k_B T_\alpha}(\mathbf{v}-\mathbf{u})^2\right]^{\frac{1}{1-q_\alpha}-1}$$

$$= \left(\frac{\partial u_x}{\partial z} + \frac{\partial u_z}{\partial x}\right)(2\pi)^{\frac{3}{2}}\left[\frac{m_\alpha(q_\alpha-1)}{k_B T_\alpha}\right]^{-\frac{7}{2}} \frac{\Gamma\left(\frac{1}{q_\alpha-1}-\frac{5}{2}\right)}{\Gamma\left(\frac{q_\alpha}{q_\alpha-1}\right)}, \quad 1 < q_\alpha < \frac{7}{5}. \tag{A.2}$$

Substituting (A.2) into (A.1) we can obtain Eq.(20).

For $q_\alpha < 1$, the integral in (A.1) is calculated by

$$\int d\mathbf{v}(v_x - u_x)(v_z - u_z)A(\mathbf{r},\mathbf{v})\left[1 - \frac{(1-q_\alpha)m_\alpha}{2k_B T_\alpha}(\mathbf{v}-\mathbf{u})^2\right]^{\frac{1}{1-q_\alpha}-1}$$

$$= \int_{-a}^{a} dv_y \int_{-b}^{b} dv_x (v_x - u_x) \int_{-c}^{c} dv_z (v_z - u_z) A(\mathbf{r},\mathbf{v})\left[1 - \frac{(1-q_\alpha)m_\alpha}{2k_B T_\alpha}(\mathbf{v}-\mathbf{u})^2\right]^{\frac{1}{1-q_\alpha}-1}, \tag{A.3}$$

where we denote

$$a = \sqrt{\frac{2k_B T_\alpha}{(1-q_\alpha)m_\alpha}}, \quad b = \sqrt{\frac{2k_B T_\alpha}{(1-q_\alpha)m_\alpha} - (v_y - u_y)^2}, \quad c = \sqrt{\frac{2k_B T_\alpha}{(1-q_\alpha)m_\alpha} - (v_y - u_y)^2 - (v_x - u_x)^2}.$$

So (A.3) can be reduced to

$$\left(\frac{\partial u_x}{\partial z} + \frac{\partial u_z}{\partial x}\right)\int_{-a}^{a} dv_y \int_{-b}^{b} dv_x (v_x - u_x)^2 \int_{-c}^{c} dv_z (v_z - u_z)^2 \left[1 - \frac{(1-q_\alpha)m_\alpha}{2k_B T_\alpha}(\mathbf{v}-\mathbf{u})^2\right]^{\frac{1}{1-q_\alpha}-1}$$

$$= -\left(\frac{\partial u_x}{\partial z} + \frac{\partial u_z}{\partial x}\right)\frac{(2\pi)^{\frac{3}{2}}(m_\alpha/k_B T_\alpha)^{-\frac{7}{2}} \Gamma\left(\frac{1}{1-q_\alpha}+3\right)}{(q_\alpha-2)(2q_\alpha-3)(1-q_\alpha)^{\frac{1}{2}} \Gamma\left(\frac{1}{1-q_\alpha}+\frac{7}{2}\right)}. \tag{A.4}$$

Substituting (A.4) into (A.1) we can obtain Eq.(20).

**APPENDIX B**

For $q_\alpha > 1$, Eq.(29) is

$$-\frac{2\pi n_\alpha m_\alpha}{\nu_\alpha} B_{q,\alpha} \left(\frac{m_\alpha}{2\pi k_B T_\alpha}\right)^{\frac{5}{2}} \int d\mathbf{v}(v_x - u_x)^2 A(\mathbf{r},\mathbf{v})\left[1 - \frac{(1-q_\alpha)m_\alpha}{2k_B T_\alpha}(\mathbf{v}-\mathbf{u})^2\right]^{\frac{1}{1-q_\alpha}-1}$$

$$= -\frac{2\pi n_\alpha m_\alpha}{\nu_\alpha} B_{q,\alpha} \left(\frac{m_\alpha}{2\pi k_B T_\alpha}\right)^{\frac{5}{2}} C(\mathbf{r},\mathbf{v}) \int_0^{+\infty} dv v^6 \left[1 - \frac{(1-q_\alpha)m_\alpha}{2k_B T_\alpha}(\mathbf{v}-\mathbf{u})^2\right]^{\frac{1}{1-q_\alpha}-1}, \tag{B.1}$$

where

$$C(r,v) = \frac{\partial u_x}{\partial x}\int_0^\pi d\theta \sin^4\theta \int_0^{2\pi} d\varphi \cos^4\varphi + \frac{\partial u_y}{\partial y}\int_0^\pi d\theta \sin^4\theta \int_0^{2\pi} d\varphi \cos^2\varphi \sin^2\varphi$$

$$+ \frac{\partial u_z}{\partial z}\int_0^\pi d\theta \sin^2\theta \cos^2\theta \int_0^{2\pi} d\varphi \cos^2\varphi$$

$$= (2\pi)^{\frac{3}{2}}\left(3\frac{\partial u_x}{\partial x} + \frac{\partial u_y}{\partial y} + \frac{\partial u_z}{\partial z}\right), \tag{B.2}$$

and

$$\int_0^{+\infty} dv\, v^6 \left[1 - \frac{(1-q_\alpha)m_\alpha}{2k_B T_\alpha}(\mathbf{v}-\mathbf{u})^2\right]^{\frac{1}{1-q_\alpha}-1} = \left[\frac{m_\alpha(q_\alpha-1)}{k_B T_\alpha}\right]^{-\frac{7}{2}} \frac{\Gamma\left(\frac{1}{q_\alpha-1}-\frac{5}{2}\right)}{\Gamma\left(\frac{q_\alpha}{q_\alpha-1}\right)}, \quad 1 < q_\alpha < \frac{7}{5}.$$

Thus Eq.(29) is equal to

$$-\frac{2n_\alpha k_B T_\alpha}{\nu_\alpha(7-5q_\alpha)}\left(3\frac{\partial u_x}{\partial x} + \frac{\partial u_y}{\partial y} + \frac{\partial u_z}{\partial z}\right), \quad 1 < q_\alpha < \frac{7}{5}, \tag{B.3}$$

For $q_\alpha < 1$, the integral in Eq.(29) is

$$\int d\mathbf{v}(v_x-u_x)^2 A(\mathbf{r},\mathbf{v})\left[1-(1-q_\alpha)\frac{m_\alpha(\mathbf{v}-\mathbf{u})^2}{2k_B T_\alpha}\right]^{\frac{1}{1-q_\alpha}-1}$$

$$= \int_{-a}^{a} dv_z \int_{-b}^{b} dv_y \int_{-c}^{c} dv_x (v_x-u_x)^2 A(\mathbf{r},\mathbf{v})\left[1-(1-q_\alpha)\frac{m_\alpha(\mathbf{v}-\mathbf{u})^2}{2k_B T_\alpha}\right]^{\frac{1}{1-q_\alpha}-1}, \tag{B.4}$$

where we have denoted that

$$a = \sqrt{\frac{2k_B T_\alpha}{(1-q_\alpha)m_\alpha}}, \quad b = \sqrt{\frac{2k_B T_\alpha}{(1-q_\alpha)m_\alpha}-(v_z-u_z)^2}, \quad c = \sqrt{\frac{2k_B T_\alpha}{(1-q_\alpha)m_\alpha}-(v_y-u_y)^2-(v_z-u_z)^2}$$

So Eq.(29) becomes

$$-\frac{2\pi n_\alpha m_\alpha}{\nu_\alpha} B_{q,\alpha}\left(\frac{m_\alpha}{2\pi k_B T_\alpha}\right)^{\frac{5}{2}} \left\{\frac{\partial u_x}{\partial x}\int_{-a}^{a} dv_z \int_{-b}^{b} dv_y \int_{-c}^{c} dv_x (v_x-u_x)^4 \left[1-\frac{(1-q_\alpha)m_\alpha}{2k_B T_\alpha}(\mathbf{v}-\mathbf{u})^2\right]^{\frac{1}{1-q_\alpha}-1}\right.$$

$$+\frac{\partial u_y}{\partial y}\int_{-a}^{a} dv_z \int_{-b}^{b} (v_y-u_y)^2 dv_y \int_{-c}^{c} dv_x (v_x-u_x)^2 \left[1-\frac{(1-q_\alpha)m_\alpha}{2k_B T_\alpha}(\mathbf{v}-\mathbf{u})^2\right]^{\frac{1}{1-q_\alpha}-1}$$

$$\left.+\frac{\partial u_z}{\partial z}\int_{-a}^{a} (v_z-u_z)^2 dv_z \int_{-b}^{b} dv_y \int_{-c}^{c} dv_x (v_x-u_x)^2 \left[1-\frac{(1-q_\alpha)m_\alpha}{2k_B T_\alpha}(\mathbf{v}-\mathbf{u})^2\right]^{\frac{1}{1-q_\alpha}-1}\right\}$$

$$= -\frac{n_\alpha k_B T_\alpha}{\nu_\alpha} B_{q,\alpha} \frac{(1-q_\alpha)^{-\frac{1}{2}}\Gamma\left(\frac{1}{1-q_\alpha}+3\right)}{(q_\alpha-2)(2q_\alpha-3)\Gamma\left(\frac{1}{1-q_\alpha}+\frac{7}{2}\right)}\left(3\frac{\partial u_x}{\partial x} + \frac{\partial u_y}{\partial y} + \frac{\partial u_z}{\partial z}\right)$$

$$= -\frac{2n_\alpha k_B T_\alpha}{\nu_\alpha(7-5q_\alpha)}\left(3\frac{\partial u_x}{\partial x} + \frac{\partial u_y}{\partial y} + \frac{\partial u_z}{\partial z}\right), \quad q_\alpha < 1. \tag{B.5}$$

This is Eq.(29).